# Ultrafast and long-term stability Integrated Pockels laser with thin-film PZT

Yueyang Zhang[1+], Chenlei Li[1+], Tao Shu[1], Wei Chen[1], Cunyu Shi[1], Feng Qiu[3], and Daoxin Dai[1,2*]

[1]State Key Laboratory for Extreme Photonics and Instrumentation, Zhejiang Key Laboratory of Optoelectronic Information Technology, College of Optical Science and Engineering, International Research Center for Advanced Photonics, Zhejiang University, Zijingang Campus, Hangzhou 310058, China;

[2]China Jiliang University, Hangzhou 310018, China;

[3]Hangzhou Institute for Advanced Study, University of Chinese Academy of Sciences, Hangzhou 310024, Zhejiang Province, China;

*Corresponding Author: *dxdai@zju.edu.cn*

[+]*These authors contributed equally*

**Abstract**: Integrated tunable lasers are central to coherent communications, wavelength-routed optical interconnects, spectroscopy and frequency-modulated continuous-wave LiDAR, yet chip-scale sources rarely combine broad wavelength coverage, nanosecond switching, high spectral purity and stable high-power operation. Here we demonstrate a frequency-agile hybrid external-cavity laser enabled by the Pockels effect in thin-film lead zirconate titanate (PZT). The strong linear electro-optic response of PZT provides direct, non-thermal tuning of compact microring resonators with a wavelength-tuning efficiency of 17 pm/V. In contrast to conventional anisotropic Pockels materials, the near-isotropic in-plane electro-optic behaviour of thin-film PZT relaxes crystal-axis layout constraints, allowing efficient Vernier wavelength selection in compact ring cavities. The PZT resonators also show no measurable photorefractive resonance distortion and no resolvable DC-bias drift during operation, preserving stable wavelength-selective feedback. The demonstrated laser achieves an 82 nm tuning range, a 5 mW fiber-coupled output power, a side-mode suppression ratio (SMSR) exceeding 56.7 dB, and a wavelength-switching time of 5.5 ns. These results establish thin-film PZT photonics as a powerful electro-optic platform for compact, high-power, and frequency-agile integrated laser sources.

## Introduction

Tunable lasers are key light sources for a wide range of photonic systems, including optical communications [1,2], data-center interconnects[3,4], spectroscopy [5,6] and frequency-modulated continuous-wave LiDAR [7-10]. As these systems move towards higher bandwidth, lower latency and greater functional reconfigurability, the laser source is no longer required only to provide a narrow-linewidth carrier. It must also access a broad wavelength range, switch rapidly between wavelength channels, maintain stable single-mode operation and deliver sufficient optical power for system-level use. These requirements are especially important for frequency-agile photonic systems, such as wavelength-routed optical interconnects[11,12], optical packet switching[13,14], wavelength-steered LiDAR[7,15] and rapid multi-channel spectral interrogation[16,17], where slow thermal wavelength scanning is insufficient and nanosecond-scale random access to selected wavelengths is highly desirable.

With the booming in photonic integrated circuits, integrated external-cavity lasers (ECLs) have emerged as a powerful alternative by combining a III-V gain section with a reconfigurable wavelength-selective external cavity [18-28]. This architecture decouples optical gain from spectral control, thereby enabling enhanced coherence, expanded tunability and greater design flexibility. Chip-scale integrated tunable lasers are therefore critical not only for improving system performance, but also for achieving practical scalability and manufacturability in large-scale photonic platforms. However, realizing such a source in an integrated chip-scale format remains challenging because the physical mechanisms used for wavelength tuning often impose conflicting constraints on tuning range, switching speed, cavity footprint, spectral stability and output power.

Recent advances across multiple integrated photonic platforms have produced ECLs with outstanding performance. Silicon and silicon nitride platforms [20-28], for example, provide highly manufacturable and low-loss external cavities. $Si_3N_4$ based lasers have achieved fiber-laser-level coherence, intrinsic linewidths in the sub-10-Hz regime [24], and wavelength tuning ranges beyond 170 nm [25]. However, wavelength reconfiguration in these low-loss Si and $Si_3N_4$ ECLs is typically implemented through thermo-optic tuning, whose response is limited by thermal diffusion and therefore occurs on microsecond timescales. For example, previously reported thermally tuned Si and $Si_3N_4$ ECLs showed wavelength-switching times of approximately 15 μs and 200 μs[26,27], respectively. Even with faster stress-optic actuation, the switching time of $Si_3N_4$-based ECLs remains in the microsecond regime[28]. Therefore, although silicon and $Si_3N_4$ based ECLs excel in tunability and low propagation loss , they are less suited to truly high-speed wavelength-reconfigurable operation.

Whereas, EO platforms such as thin-film lithium niobate (TFLN) enable genuinely fast cavity reconfiguration [29-32]. For example, in Ref. [29], a hybrid $Si_3N_4$-TFLN laser achieved a 40 nm tuning range together with sub-50-ns switching over 12 nm. Nevertheless, large EO wavelength shifts remain difficult to realize in compact TFLN resonant cavities [33]. Owing to its moderate electro-optic coefficient ($r_{33} \approx 30$ pm $V^{-1}$ )[34] and strong anisotropy, TFLN requires careful alignment of the optical polarization, applied electric field, and crystal axis to achieve efficient EO modulation. As a result, resonant tuners on TFLN are typically implemented as racetrack cavities containing long straight waveguide segments aligned along the extraordinary axis. Because of the limited EO coefficient, these straight sections must be extended to accumulate sufficient phase shift, leading to exceptionally large resonator circumferences and consequently increased device footprint. For instance, in Ref. [29], the racetrack resonator had a free spectral range of only ~51 GHz and was described as a large-circumference resonator with an elongated EO-tuning section, which enhances EO efficiency but comes at the expense of FSR and optical loss . Beyond this geometric trade-off, LN resonators can suffer from DC bias drift under sustained electrical operation [35]. At high intracavity power, PR effects can induce long-lived refractive-index changes that cause resonance drift and lineshape distortion, destabilizing the wavelength-selective feedback in an ECL and leading to lasing-frequency drift, output-power fluctuation and a limited stable output-power range [36,37]. Therefore, although TFLN excels in high-speed EO reconfiguration, a compact platform that simultaneously offers stronger EO tuning efficiency and improved operational stability remains highly desirable.

Recently, ferroelectric thin-film lead zirconate titanate (PZT) offers an attractive alternative for integrated tunable lasers. In addition to its large EO response [38], thin-film PZT exhibits a near-isotropic in-plane EO behavior [39], which relaxes cavity-layout constraints and is particularly advantageous for compact resonant photonic circuits. Compared with conventional EO materials, PZT also shows negligible susceptibility to PR effects under optical operation, thereby supporting more stable spectral behavior. Owing to these attributes, PZT integrated photonics has already enabled high-performance modulators and other actively reconfigurable devices with excellent efficiency and compactness [38,40-42]. This leaves open the possibility of a laser architecture that directly harnesses the strong EO response of PZT for rapid and efficient cavity reconfiguration.

Here, we introduce a PZT-based hybrid ECL that addresses the longstanding difficulty of combining broad wavelength tunability, high frequency agility and stable single-mode operation within a compact resonant cavity.

Our device is built on a dual-ring architecture, in which the strong EO interaction, exceptionally small DC drift and nearly negligible PR response of thin-film PZT are harnessed together with highly selective wavelength feedback from two detuned microring resonators (MRRs). Experimentally, the laser achieves an 82 nm tuning range, a minimum wavelength switching time of 5.5 ns, a linewidth of 10 kHz, a 5 mW fiber coupled output power and an SMSR exceeding 56.7 dB, with no observable DC drift during operation. To the best of our knowledge, this is the first ECL realized on a PZT photonic platform, and it establishes the widest tuning range reported for a Vernier EO-tuned ECL together with one of the fastest wavelength-switching speeds among EO-tuned ECLs. These results position PZT integrated photonics as a promising foundation for compact, frequency-agile and spectrally pure on-chip laser sources. Together, these results establish a frequency-agile hybrid ECL that integrates broad wavelength tunability, high frequency agility and stable single-mode operation, highlighting its potential for reconfigurable optical interconnects, precision spectroscopy and coherent LiDAR systems.

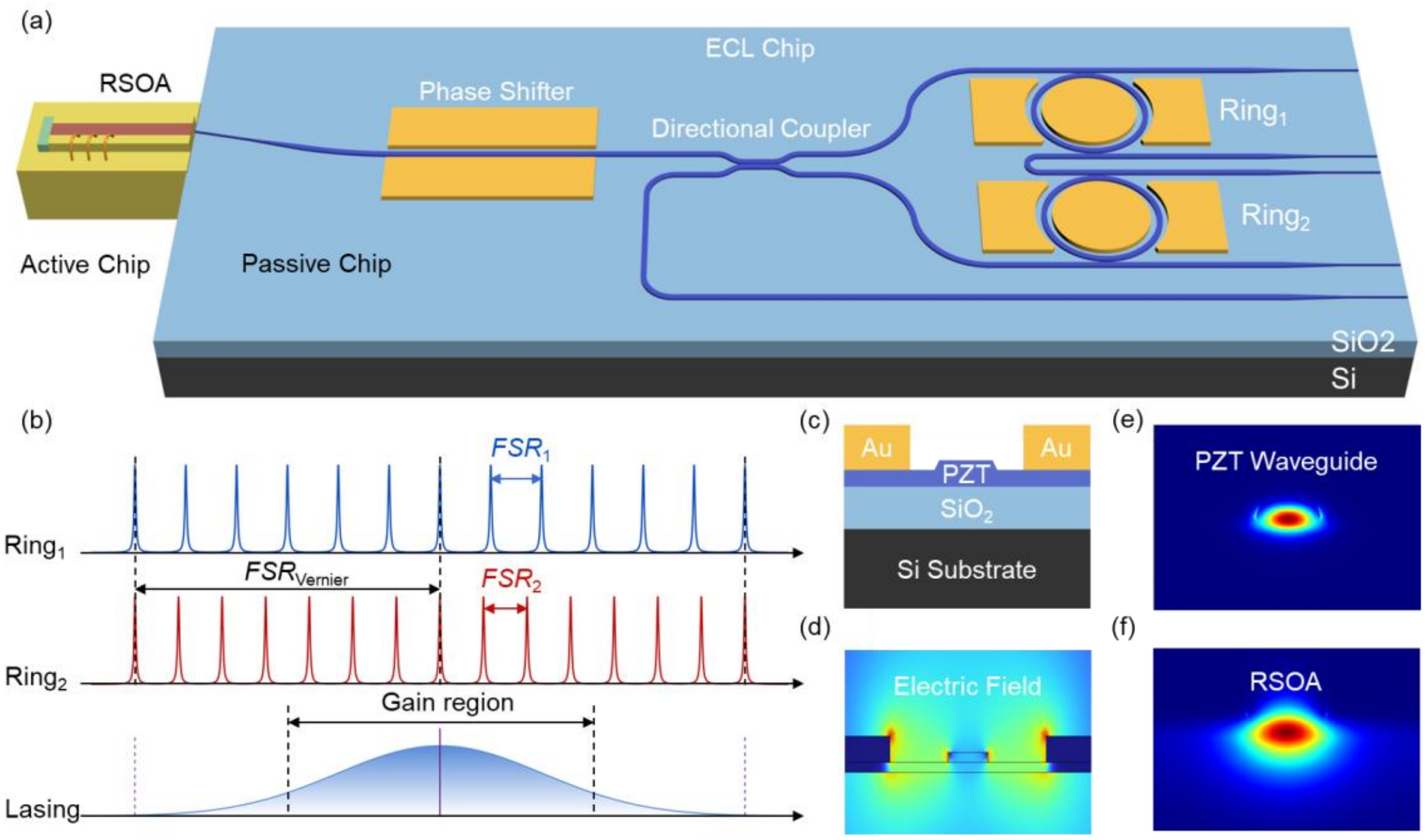

**Fig. 1** Device concept and structural design of the hybrid PZT ECL. (a) Schematic illustration of the hybrid ECL; (b) Operating principle of the dual-ring Vernier cavity; (c) Cross-sectional schematic of the PZT waveguide with lateral Au electrodes for EO tuning; (d) Electric field image of the fabricated PZT ridge waveguide; (e) Simulated optical mode profile of the PZT waveguide; (f) Simulated optical mode profile of the RSOA output waveguide.

## Results

### Laser Design and Fabrication

As illustrated in Fig. 1 (a), the proposed laser consists of a reflective semiconductor optical amplifier (RSOA) hybrid-integrated with a PZT photonic chip, where the external cavity incorporates a phase shifter, a directional coupler, and a pair of MRRs forming a Vernier-type wavelength-selective feedback circuit. This architecture decouples optical gain and wavelength control, allowing each function to be independently optimized, while maintaining a compact footprint and compatibility with scalable photonic integration.

The wavelength-selective feedback is implemented using a dual-ringarchitecture, as illustrated in Fig. 1 (b). Such configurations have been widely adopted in integrated ECLs to enhance spectral selectivity and enable extended wavelength tuning through the Vernier effect. In this scheme, two MRRs with slightly different FSRs are coupled to a common bus waveguide, and lasing occurs only at the wavelengths where their resonances overlap, thereby ensuring single-mode operation with high spectral purity.The strong and near-isotropic in-plane electro-optic behavior in thin-film PZT allows the use of compact MRR geometries while still maintaining efficient electrical control.

The EO tuning functionality is implemented within a PZT waveguide platform, and the corresponding device structure is shown in Fig. 1 (c-e). The optical mode is confined in a ridge waveguide formed in the PZT layer, while metal electrodes are positioned on both sides to generate a transverse electric field overlapping with the guided mode. This configuration enables efficient refractive index modulation through the Pockels effect, allowing direct EO control of the cavity resonance. Owing to the strong EO interaction in PZT, the electrode length can be kept relatively short, which is essential for preserving compact MRR geometries and avoiding excess optical loss. Fig. 1 (d) and (e) show the electric field distribution generated by an external bias of 1 V and the optical field profile of the $TE_0$ mode, respectively. The optical field profile of the $TE_0$ mode in the RSOA waveguide is shown in Fig. 1 (f).Together, these structural and fabrication features establish a compact and robust platform for implementing fast and stable EO tuning within the external cavity.

To rationally design the wavelength-selective reflector of the laser, we first analyze the spectral response of the dual-ring Vernier cavity. The two MRRs are designed with slightly different radii of 150 μm and 152 μm, respectively, so that their resonance conditions are intentionally detuned while remaining sufficiently close to generate a strong Vernier effect. The simulated transmission spectrum exhibits a pronounced envelope with an effective Vernier FSR of 76 nm (Figure S1 of Supplementary section 1), together with a SMSR of approximately 6 dB at the filter level. This enlarged synthetic FSR is substantially greater than that of either individual ring, and therefore provides the spectral basis for wide-range wavelength selection in the ECL, while the overlap of the two ring resonances ensures preferential feedback for the target lasing mode.

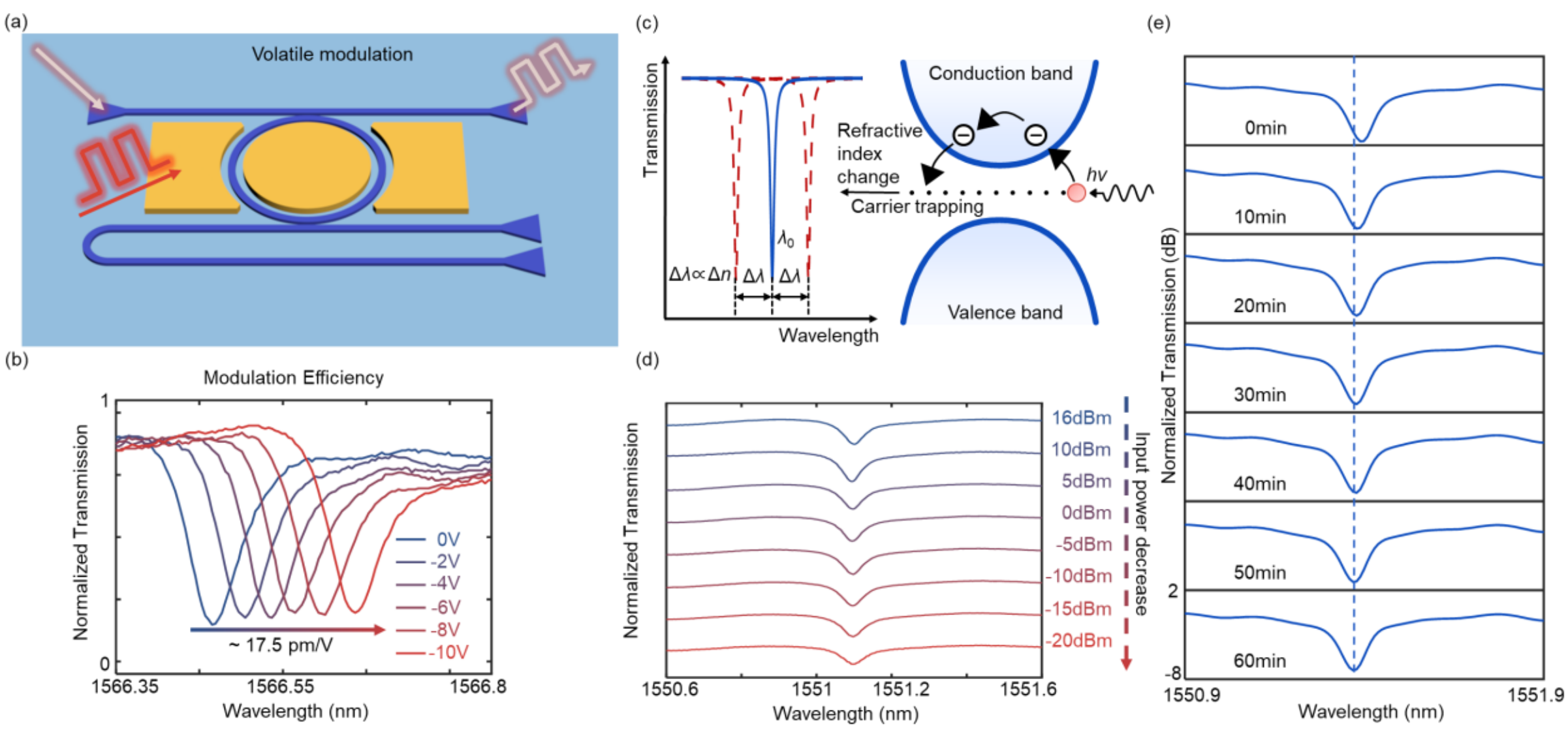

**Fig. 2** EO characterization and PR effect of the PZT resonator platform. (a) Schematic of volatile EO modulation in a single PZT MRR resonator; (b) Measured transmission spectra of the PZT MRR resonator under different applied voltages; (c) Under optical illumination, electrons trapped at impurity levels are excited into the conduction band, triggering the PR effect. ; (d) Laser-scanned transmission spectra of a PZT ring resonator under different optical power with a fixed scan rate 0.5 nm/s. The power varies from 16 dBm to -20 dBm; (e) Long-term resonance stability of the PZT resonator under 5V bias, showing negligible direct-current drift.

**Wavelength tuning and PR effect of PZT single MRR**

To evaluate the EO tuning capability of the PZT platform at the single-resonator level, we first characterized a representative PZT MRR. The MRR resonators were fabricated on a 300-nm-thick thin-film PZT layer. As schematically illustrated in Fig. 2 (a), the device works by electrically shifting the cavity resonance through the Pockels effect after domain initialization of the PZT film.

In the present experiment, the resonator was first poled to establish a stable ferroelectric state prior to EO operation. After the poling voltage was removed, the polarization of the PZT was frozen. No further application of a DC poling bias was required during EO operation. The measured transmission spectra under different driving voltages are shown in Fig. 2 (b). A clear and continuous resonance displacement is observed as the bias is varied, confirming efficient EO tuning in a compact MRR geometry. From the measured spectral shift, the EO tuning efficiency is extracted to be approximately 17.5 pm/V, which is consistent with the strong EO response previously reported for PZT resonant devices and is markedly higher than that typically obtained in thin-film lithium-niobate MRR modulators.

The PR stability of the PZT MRR resonator is further evaluated. Fig. 2 (c) schematically illustrates the PR process, in which electrons trapped at defect or impurity levels are optically excited into the conduction band, migrate through the material and are subsequently retrapped at available defect sites, thereby forming a space-charge field that modifies the refractive index through the Pockels effect.Through the Pockels effect, this field modifies the refractive index and can induce resonance drift, lineshape distortion and power instability in resonators. Such effects have been clearly observed in TFLN microrings[36], where increasing optical power leads to pronounced resonance deformation and frequency shifts, and light-induced refractive-index changes can persist for more than 10 h in certain TFLN devices.

In contrast, the PZT MRR shows no measurable PR effect within the tested power range. As shown in Fig. 2 (d), the resonance wavelength, Q factor and spectral lineshape remain nearly unchanged as the input power is increased, with no observable resonance drift or distortion. Additional measurements performed at different wavelength-scanning speeds, together with the corresponding test setup, are provided in Figure S1, S2 of Supplementary section 2. These results reveal the negligible PR susceptibility of thin-film PZT, establishing a stable resonant platform for subsequent ECL operation.

In addition to tuning efficiency and PR effect, the operational stability of the resonator under sustained bias is a key consideration for ECL, which require a well-defined cavity resonance during continuous operation. In lithium-niobate-based resonant devices, bias drift remains a persistent challenge and often necessitates continuous recalibration or active feedback. By contrast, the PZT resonator exhibits negligible DC drift. As shown in Fig. 2 (e), the resonance wavelength remains nearly unchanged over the full observation window under continuous bias, confirming highly stable EO operation. This drift-suppressed behavior is particularly beneficial for integrated ECLs, as it preserves cavity stability while enabling rapid resonance tuning.

**Static Laser Characterization**

The corresponding device implementation is shown in Fig. 3 (a). The passive cavity is realized on the PZT platform and hybrid-integrated with an L-band InP RSOA. The anti-reflection (AR)-coated facet of the RSOA is butt-coupled to the PZT PIC through an angled waveguide taper, whereas the opposite RSOA facet is high-reflectivity (HR)-coated to serve as the back mirror of the laser cavity. This hybrid configuration separates gain generation from EO wavelength selection, allowing the RSOA and the PZT cavity to be independently optimized while maintaining a compact and robust laser architecture.

We first characterize the amplified spontaneous emission (ASE) of the RSOA at different injection currents, as shown in Fig. 3 (b). As the current increases from 90 mA to 210 mA, the ASE intensity rises steadily while preserving a broad spectral envelope across the C+L band. This gain profile defines the spectral window available for the hybrid ECL and provides the basis for subsequent wavelength-selection experiments using the Vernier cavity. To balance sufficient gain with stable operation, an injection current of 120 mA was chosen for the following wavelength-tuning measurements.

As shown in Fig. 3 (a), the hybrid laser was evaluated on a precision multi-axis alignment platform with independent electrical control of the active and passive sections. During the measurements, the RSOA was mounted on a thermally regulated holder and operated at 25 ℃, so that the gain condition remained stable throughout the spectral and coherence characterization. The laser output was routed through a 9:1 fiber splitter, enabling simultaneous readout of the emitted power and optical spectrum. This arrangement allowed the lasing state to be followed continuously during cavity tuning. With the coupling condition optimized, the device produced a maximum fiber-coupled output power of about 5 mW. The corresponding light-current characteristics of the PZT-based ECL are provided in Figure S1 of Supplementary Section 3. Considering the ~6 dB coupling loss (Figure S2 of Supplementary section 3) at the fabricated fiber-coupling facet, the actual laser output power is estimated to be approximately 20 mW.

With the injection current fixed at 120 mA, we then investigated the static tuning behavior of the laser by electrically controlling the PZT cavity. A representative lasing spectrum is shown in (c), where an SMSR exceeding 56.ed, indicating strong single-mode discrimination enabled by the dual-ring Vernier cavity. The static wavelength tuning results are summarized in Fig. 3 (d). Starting from the shorter-wavelength side, the lasing wavelength was first shifted coarsely by applying voltages to the individual rings to move the dual-ring Vernier overlap across the gain window, while the integrated phase shifter was simultaneously adjusted to compensate for the cavity phase and maintain mode alignment. In the present experiment, the tuning voltages applied to the PZT cavity were varied from 0 V to 30 V. By jointly controlling the two ring resonators and the phase shifter, the laser wavelength could be displaced continuously from 1550 nm to 1632 nm, corresponding to an overall tuning range of 82 nm. Throughout the tuning process, single-mode lasing was preserved across the full spectral span, demonstrating that the PZT-based dual-ring cavity can simultaneously provide broad EO wavelength reconfiguration and strong spectral selectivity in a compact ECL architecture.

The long-term spectral stability of the hybrid ECL was further evaluated by monitoring the lasing wavelength under sustained electrical bias. In this measurement, a constant voltage of 5 V was continuously applied to one of the PZT rings, and the lasing spectrum was recorded over a period of 1 h. Remarkably, no observable wavelength drift was detected throughout the full measurement window in Fig. 3 (e). This result indicates that the drift-suppressed behavior of the PZT resonator is preserved after incorporation into the complete laser cavity, allowing the resonant condition to remain stable even under continuous electrical operation. Such stability is particularly important for EO-tuned ECLs, where even small cavity fluctuations can perturb the lasing wavelength and compromise repeatability during tuning or switching.

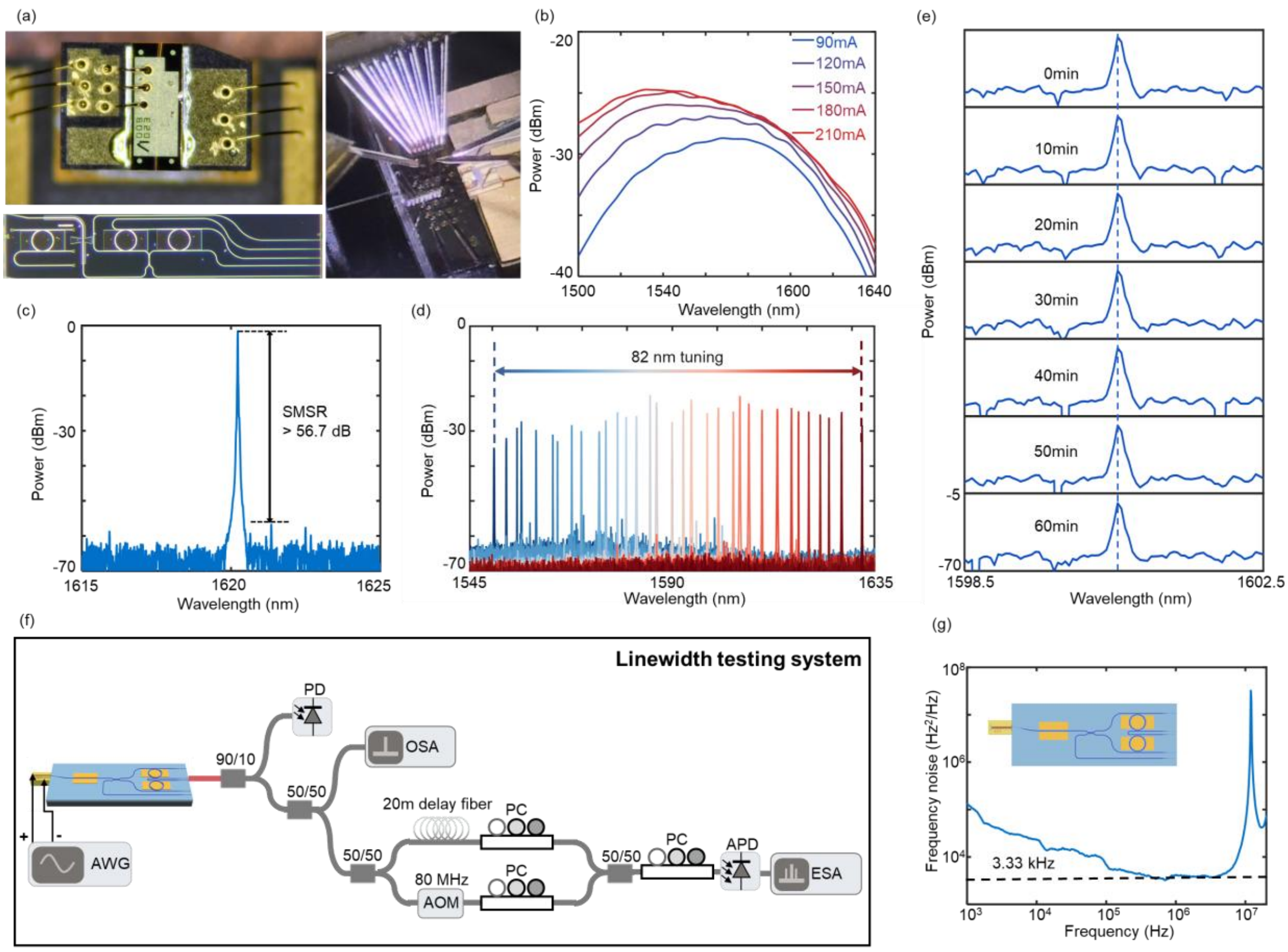

**Fig. 3** Static characterization of the hybrid PZT external-cavity laser. (a) Photograph of the RSOA , the fabricated hybrid device under optical testing; (b) Amplified spontaneous emission spectra of the RSOA measured at different injection currents; (c) High-resolution lasing spectrum of the hybrid ECL; (d) Superimposed lasing spectra during static wavelength tuning; (e) Long-term lasing-wavelength stability under sustained electrical bias; (f) Experimental setup for linewidth characterization based on phase-noise measurement; (g) Measured phase-noise spectrum used to extract the laser linewidth.

The laser coherence was characterized using the phase-noise measurement chain shown in Fig. 3 (f). In this setup, a small fraction of the laser output was first tapped for real-time power monitoring, while the remaining signal was directed to the linewidth-characterization branch. The optical field was then split into two paths: one propagated through a 20m fiber delay line, whereas the other passed through an acousto-optic modulator to introduce a frequency shift. After recombination and polarization adjustment, the delayed beat signal was detected and analyzed on an electrical spectrum analyzer (ESA). In this configuration, the fluctuations of the instantaneous laser frequency are converted into fluctuations of the interferometer output phase, from which the laser phase-noise spectrum can be obtained. This measurement principle is analogous to the delayed self-heterodyne phase-noise characterization commonly used for narrow-linewidth hybrid lasers, where the phase fluctuations of the delayed beat note are first measured and then converted to frequency-noise information. The detailed measurement principle and noise-conversion procedure are provided in Supplementary section 4.

From the measured spectrum in Fig. 3 (g), the white-noise floor is extracted as 3.33 kHz, corresponding to a laser linewidth of 10.46 kHz. This result confirms that the present hybrid PZT ECL preserves kilohertz-level spectral coherence while simultaneously supporting wide EO tunability and negligible DC drift.

**Ultrafast wavelength switching**

High-speed wavelength switching is a key capability for reconfigurable photonic systems. In wavelength-routed optical switching and optical packet-switched networks [11-13], nanosecond-scale access to different wavelength channels is required to reduce packet loss and increase network throughput. Such operation is also relevant to data-center optical circuit switching [2,14], where rapidly tunable laser sources can support dynamic wavelength assignment and low-latency path reconfiguration. Beyond communications, fast wavelength hopping can enable wavelength-steered LiDAR [7,15], in which different lasing wavelengths address different emission angles through dispersive photonic antennas, as well as rapid interrogation of multi-channel optical sensors and spectroscopic signatures. The nanosecond-scale switching demonstrated here therefore provides a route towards on-chip laser sources for wavelength-routed interconnects, solid-state LiDAR and reconfigurable sensing systems.

Leveraging the strong EO response of the PZT dual-ring cavity, we next demonstrate ultrafast wavelength switching of the hybrid ECL. The two targeted lasing modes are located at 1605.57 nm and 1615.42 nm, respectively, corresponding to a wavelength spacing of 9.85 nm. Under electrical driving, the EO-induced resonance shift modifies the Vernier overlap condition of the dual-ring cavity and thereby forces the laser to switch deterministically between these two discrete lasing wavelengths.

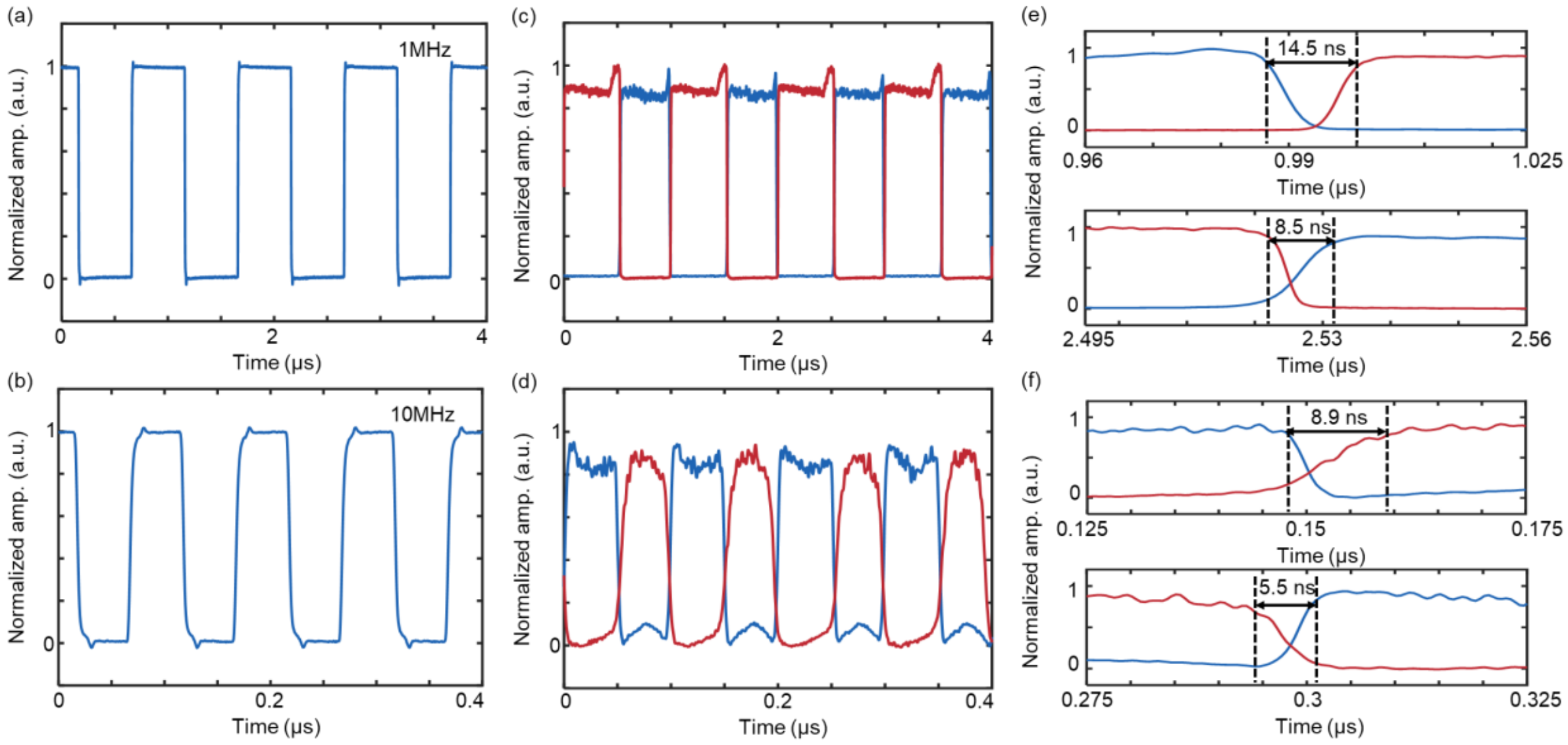

**Fig. 4** Ultrafast wavelength switching of the hybrid PZT external-cavity laser. (a,b), Applied square-wave electrical signals at modulation frequencies of 1 MHz and 10 MHz; (c,d) Corresponding time-domain optical switching traces between the two lasing modes at 1 MHz and 10 MHz; (e) Enlarged views of the switching transients at 1 MHz, yielding wavelength switching times of 14.5 ns and 8.5 ns; (f) Enlarged views of the switching transients at 10 MHz, yielding wavelength switching times of 8.9 ns and 5.5 ns.

The measured electrical driving waveforms at 1 MHz and 10 MHz are shown in Fig. 4 (a) and (b), respectively, confirming that the packaged EO actuator can sustain high-speed square-wave operation. The corresponding optical switching traces are presented in Fig. 4 (c) and (d), where repeatable transitions between the two lasing states are clearly observed at both modulation frequencies. Compared with the 1 MHz case, the 10 MHz traces exhibit slightly increased distortion and residual fluctuation, which is attributed to the finite cavity-settling dynamics under faster triggering. Nevertheless, the switching remains robust and well resolved, demonstrating that the PZT ECL can maintain deterministic wavelength reconfiguration at multi-megahertz rates.

The switching transients are quantified in Fig. 4 (e) and (f). Here, the wavelength switching time is defined as the time interval from the 90% falling edge of one wavelength state to the 90% rising edge of the other wavelength state, consistent with the convention adopted in previous integrated EO laser-switching studies. At 1 MHz, the measured switching times are 14.5 ns and 8.5 ns for the two opposite transition directions. When the modulation frequency is increased to 10 MHz, the switching remains clearly resolvable, with transition times of 8.9 ns and 5.5 ns, respectively. The shortest wavelength switching time of 5.5 ns demonstrates that the present PZT-based hybrid ECL can support nanosecond-scale switching over a spectral interval approaching 10 nm in a compact dual-ring architecture. Such performance directly benefits from the efficient EO tuning of the PZT cavity together with the compact MRR geometry, which enables rapid reconfiguration of the Vernier-selected lasing mode without relying on slow thermo-optic tuning.

Overall, these results show that the proposed PZT ECL combines broad spectral tunability with fast and repeatable EO wavelength switching. The demonstrated 9.85 nm wavelength hop and 5.5 ns minimum switching time establish the PZT dual-ring platform as a promising route toward frequency-agile on-chip laser sources for applications requiring both wide spectral coverage and high reconfiguration speed.

Table I. Comparison of wavelength swithing chip-scale lasers.

| Ref. | Platform | Mechanism | SMSR (dB) | Tuning range (nm) | Rise/Fall time (ns) | Wavelength switching time (ns) |
|---|---|---|---|---|---|---|
| **[45]** | Ⅲ-Ⅴ | SFPC | 35 | 17.76 | \ | 5 |
| **[46]** | Ⅲ-Ⅴ | SG-DBR | \ | 24.9 | \ | 5 |
| **[47]** | **Silicon** | Triple MZIs | 37 | 56 | \ | 10 |
| **[44]** | **$Si_3N_4$** | Double MRRs | 50 | 3.2 | 6 | \ |
| **[28]** | **$Si_3N_4$** | Double MRRs | 52 | 46.16 | \ | 3860 |
| **[48]** | **$LiNbO_3$** | Triple MRRs | 52 | 96 | \ | 18 |
| **[29]** | **$LiNbO_3$+ $Si_3N_4$** | Double MRRs | 63.9 | 40 | \ | 16 |
| **[30]** | **$LiNbO_3$** | Double MRRs | 50 | 20 | 3 | \ |
| **This** | **PZT** | Double MRRs | 56.7 | 82 | 3 | 5.5 |

SFPC:Slotted Fabry–Pérot cavity.

## Discussion and Conclusion

The results presented here demonstrate that a PZT-based hybrid ECL can unify a set of laser performance metrics that are difficult to achieve simultaneously in compact wavelength-agile sources. As summarized in Table I, previously reported chip-scale wavelength-switching lasers generally involve trade-offs among tuning range, switching speed and side-mode suppression. Monolithic III–V tunable lasers can reach nanosecond-scale switching, but their active semiconductor cavities typically provide limited cavity design flexibility and are susceptible to gain-index coupling, power fluctuation and mode competition during wavelength tuning. Silicon and $Si_3N_4$ external-cavity lasers offer excellent coherence and broad static tunability, yet their wavelength reconfiguration is usually limited by slow thermo-optic or stress-optic actuation. LN-based EO lasers provide a route to fast cavity tuning, but compact resonant implementations remain constrained by the moderate EO coefficient, anisotropic electrode layout, long racetrack tuning sections, DC bias drift and photorefractive perturbations under high intracavity power. In contrast, the PZT ECL demonstrated here combines an 82 nm tuning range, a minimum wavelength switching time of 5.5 ns, an SMSR of 56.7 dB, a 10 kHz linewidth and negligible DC drift, thereby defining a distinct performance regime for compact EO-tuned integrated lasers.

This performance is enabled by the material properties of thin-film PZT and their compatibility with compact resonant wavelength-selective cavities. The strong EO response and near-isotropic in-plane EO behavior of PZT allows efficient resonance control without requiring long straight tuning sections, reducing the footprint penalty typically encountered in other EO platforms. Its near-isotropic in-plane EO behavior further relaxes the cavity-layout constraints imposed by crystal-axis alignment, allowing the EO tuning function to be implemented directly in compact microring geometries. Together with Vernier-enhanced wavelength selection, this efficient resonance tuning is converted into a broad accessible spectral window while maintaining strong single-mode discrimination. Equally important, the negligible DC drift observed at both the resonator level and the complete laser level ensures reproducible cavity states under sustained electrical operation. The absence of measurable PR resonance distortion over the tested optical-power range further supports stable wavelength-selective feedback, which is essential for practical high-speed wavelength-switching lasers.

In summary, a hybrid ECL based on thin-film PZT photonics has been demonstrated, showing broad wavelength tuning, nanosecond-scale frequency agility, high side-mode suppression and stable single-mode operation in a integrated source. To the best of our knowledge, this represents the first ECL realized on a PZT photonic platform and the widest tuning range reported for a dual-ring EO-tuned ECL, together with one of the fastest wavelength switching speeds among chip-scale EO-tuned lasers. Beyond the specific device demonstrated here, this work establishes thin-film PZT as a powerful EO photonic platform for frequency-agile integrated light sources, where strong Pockels tuning, high frequency agility can be combined for reconfigurable optical interconnects, precision spectroscopy and coherent LiDAR systems.

**Acknowledgements**

We acknowledge support from the National Natural Science Foundation of China (62550180, 62405271, U23B2047, 62321166651), Fundamental and Interdisciplinary Disciplines Breakthrough Plan of the Ministry of Education of China (JYB2025XDXM106), Natural Science Foundation of Zhejiang Province (LD19F050001), Zhejiang Provincial Major Research and Development Program (2022C01103), Fundamental Research Funds for the Central Universities, and the "Pioneer" and "Leading Goose" R&D Program of Zhejiang Province (2024C01112).

**Disclosures**

The authors declare no conflicts of interest.

**Data availability**

Data underlying the results presented in this paper are not publicly available at this time but may be obtained from the authors upon reasonable request.